\documentclass[a4paper,11pt]{article}
\usepackage{jcappub} 

\usepackage{multirow}
\usepackage{soul}
\usepackage{comment}

\usepackage{hyperref}
\usepackage{graphicx}
\usepackage{subcaption}
\usepackage{comment}

\title{\boldmath First Searches for Dark Matter with the KM3NeT Neutrino Telescopes }

\collaboration{\includegraphics[height=17mm]{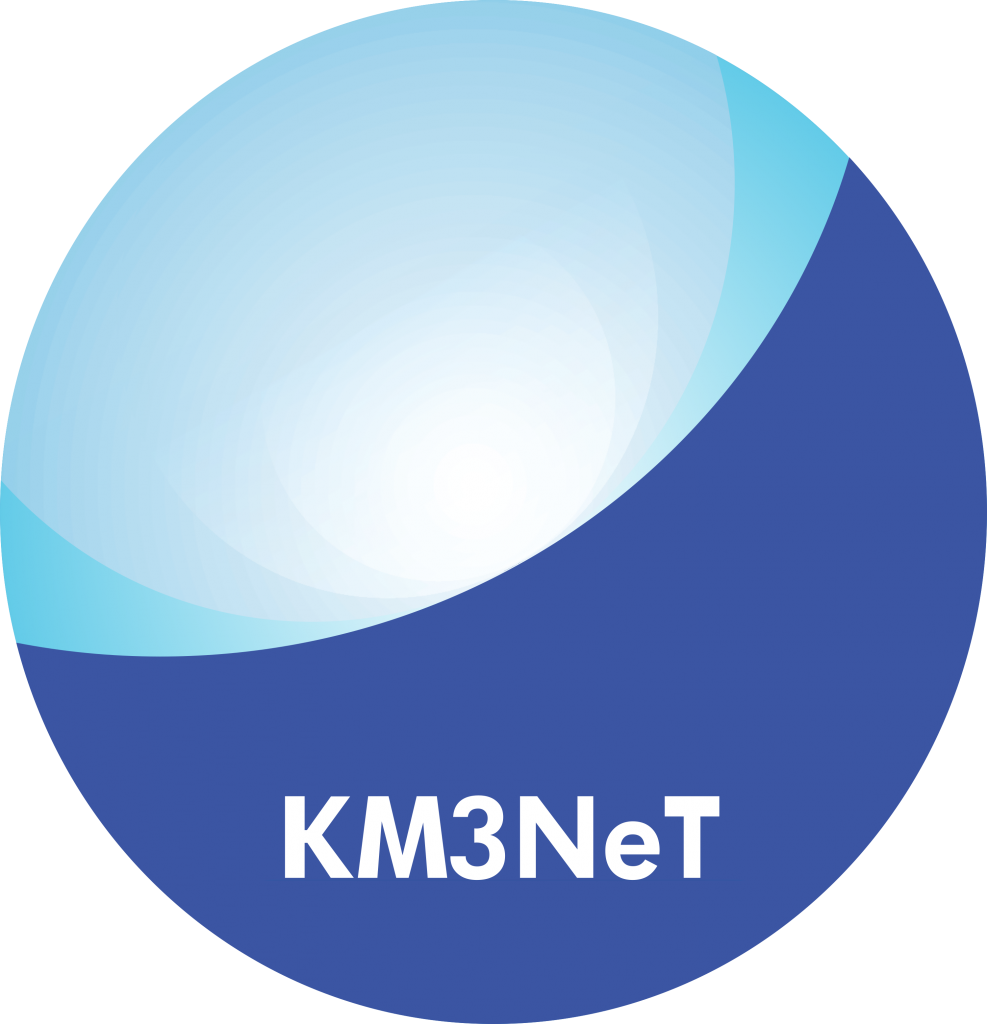}\\[6pt]
  The KM3NeT Collaboration}

\author[a]{S.~Aiello}
\author[b,bb]{A.~Albert}
\author[c]{A.\,R.~Alhebsi}
\author[d]{M.~Alshamsi}
\author[e]{S. Alves Garre}
\author[g,f]{A. Ambrosone}
\author[h]{F.~Ameli}
\author[i]{M.~Andre}
\author[j]{L.~Aphecetche}
\author[k]{M. Ardid}
% ...
\affiliation[a]{INFN, Sezione di Catania, (INFN-CT) Via Santa Sofia 64, Catania, 95123 Italy}
\affiliation[b]{Universit{\'e}~de~Strasbourg,~CNRS,~IPHC~UMR~7178,~F-67000~Strasbourg,~France}
\affiliation[c]{Khalifa University, Department of Physics, PO Box 127788, Abu Dhabi, 0000 United Arab Emirates}
\affiliation[d]{Aix~Marseille~Univ,~CNRS/IN2P3,~CPPM,~Marseille,~France}
\affiliation[e]{IFIC - Instituto de F{\'\i}sica Corpuscular (CSIC - Universitat de Val{\`e}ncia), c/Catedr{\'a}tico Jos{\'e} Beltr{\'a}n, 2, 46980 Paterna, Valencia, Spain}
\affiliation[f]{INFN, Sezione di Napoli, Complesso Universitario di Monte S. Angelo, Via Cintia ed. G, Napoli, 80126 Italy}
\affiliation[g]{Universit{\`a} di Napoli ``Federico II'', Dip. Scienze Fisiche ``E. Pancini'', Complesso Universitario di Monte S. Angelo, Via Cintia ed. G, Napoli, 80126 Italy}
\affiliation[h]{INFN, Sezione di Roma, Piazzale Aldo Moro 2, Roma, 00185 Italy}
\affiliation[i]{Universitat Polit{\`e}cnica de Catalunya, Laboratori d'Aplicacions Bioac{\'u}stiques, Centre Tecnol{\`o}gic de Vilanova i la Geltr{\'u}, Avda. Rambla Exposici{\'o}, s/n, Vilanova i la Geltr{\'u}, 08800 Spain}
\affiliation[j]{Subatech, IMT Atlantique, IN2P3-CNRS, Nantes Universit{\'e}, 4 rue Alfred Kastler - La Chantrerie, Nantes, BP 20722 44307 France}
\affiliation[k]{Universitat Polit{\`e}cnica de Val{\`e}ncia, Instituto de Investigaci{\'o}n para la Gesti{\'o}n Integrada de las Zonas Costeras, C/ Paranimf, 1, Gandia, 46730 Spain}
% ...
\affiliation[bb]{Universit{\'e} de Haute Alsace, rue des Fr{\`e}res Lumi{\`e}re, 68093 Mulhouse Cedex, France}
% ...

\emailAdd{asaina@km3net.de}
\emailAdd{mgg@ugr.es}
\emailAdd{km3net-pc@km3net.de}

\vspace{\fill}
\newpage

\abstract{
Indirect dark matter detection methods are used to observe the products of dark matter annihilations or decays originating from astrophysical objects where large amounts of dark matter are thought to accumulate.
With neutrino telescopes,  an excess of neutrinos is searched for in nearby dark matter reservoirs, such as the Sun and the Galactic Centre, which could potentially produce a sizeable flux of Standard Model particles.

The KM3NeT infrastructure, currently under construction, comprises the ARCA and ORCA undersea Čerenkov neutrino detectors located at two different sites in the Mediterranean Sea,
offshore of Italy and France, respectively. The two detector configurations are optimised for
the detection of neutrinos of different energies, enabling the search for
dark matter particles with masses ranging from a few GeV/c$^2$ to hundreds of TeV/c$^2$. In this work, searches for dark matter annihilations in the Galactic Centre and the
Sun with data samples taken with the first configurations of both detectors are presented. No significant excess over the expected background was found in either of the two analyses. Limits on the velocity-averaged self-annihilation cross section of dark matter particles are computed for five different primary annihilation channels in the Galactic Centre. For the Sun, limits on the spin-dependent and spin-independent scattering cross sections of dark matter with nucleons are given for three annihilation channels.}

\begin{document}
\maketitle

\flushbottom

\section{Introduction}

 

The existence of dark matter has been postulated to explain the dynamics of astrophysical objects in gravitational fields and to account for the structure formation in the Universe \cite{bertone}. Assuming a particle nature for dark matter, one candidate class of particles that is in agreement with the evidences of dark matter are weakly interacting massive particles (WIMPs). WIMPs can easily be accommodated in our current cosmological model: a particle with a pair-annihilation cross section analogous to the cross section of weak interactions and a mass in the GeV$/c^2 - $TeV$/c^2$ region which reproduces, via the freeze-out mechanism, the relic abundance of dark matter inferred from the power spectrum of the cosmic microwave background. Various extensions of the Standard Model (SM) can provide a candidate WIMP particle. The candidates for the lightest supersymmetric particle of the minimal supersymmetric extension of the SM, such as the neutralino, the sneutrino or the gravitino, are dark matter candidates. Other approaches to solve the gauge hierarchy problem can also provide WIMP candidates, e.g. the introduction of extra dimensions in Kaluza-Klein theories results in a number of WIMP candidate particles \cite{kaluza_klein}. Finally, a series of minimal extensions to the SM furnish a variety of other dark matter candidates~\cite{minimalDM}.

Three detection strategies are currently employed in an attempt to observe dark matter particles: direct detection, indirect detection and production at colliders. Direct detection experiments attempt to observe the WIMP-nucleon scattering process. The energy transferred to the target nuclei in the scattering is emitted in the form of scintillation light, ionisation electrons or phonons, depending on the medium \cite{bertone}. Noble gas detectors, bubble chambers and crystal scintillator detectors are the most common approaches to induce and observe the signal of nuclear recoils \cite{LZ:2022lsv, PICO_SUN, COSINE-100}. Collider searches attempt to observe evidence of dark matter particles by searching for signatures of new physics in collision events~\cite{colliders}.

Indirect detection experiments attempt to detect SM particles produced by annihilations or decays of dark matter particles. The SM products that are searched for in different experiments include neutrinos \cite{antares_gc, ANTARES_SUN, IC, IceCube_SUN, IceCube_sun_2, Super-Kamiokande_SUN}, $\gamma$-rays \cite{hess, fermiNew, magic, veritas, hawc} and antimatter (positrons, antiprotons and antideuterons) \cite{AMS, pamela, dampe, calet}. Searches for the products of dark matter annihilations are performed by observing objects where dark matter is thought to accumulate. One such object is the Galactic Centre, as galaxy formation models predict the existence of galactic dark matter halos with very high densities at their centre \cite{nfw}.

A second possible source is the Sun, where dark matter particles of the Galactic halo scatter off nuclei in the solar medium, causing them to get trapped in the gravitational potential of the Sun and accumulate in its core \cite{sun_capture}. Given enough time, the dark matter capture and annihilation processes reach  equilibrium and a steady flux of particles is predicted to be emitted from the centre of the Sun. Neutrinos below a few TeV can escape the Sun interior and reach the Earth, while no other SM particle can do so \cite{sun_Jungman}. 

This paper is organised as follows: the neutrino flux expectation due to dark matter annihilations is described in Sec. \ref{sec:dm_signal}. The KM3NeT detector, the event simulation, propagation and reconstruction methodology and the event selection are described in Sec. \ref{sec:km3net}. In Sec. \ref{sec:method}, the analysis methods used to search for neutrinos produced by dark matter annihilations are described. The results of searches for dark matter in the Galactic Centre and the Sun with KM3NeT are reported in Sec. \ref{sec:results_gc} and Sec. \ref{sec:results_Sun}, respectively. The paper is concluded with a summary in Sec. \ref{sec:summary}.

\section{Expected neutrino flux from dark matter annihilation processes}
\label{sec:dm_signal}

\begin{figure}
    \centering
    \includegraphics[width=0.7\textwidth]{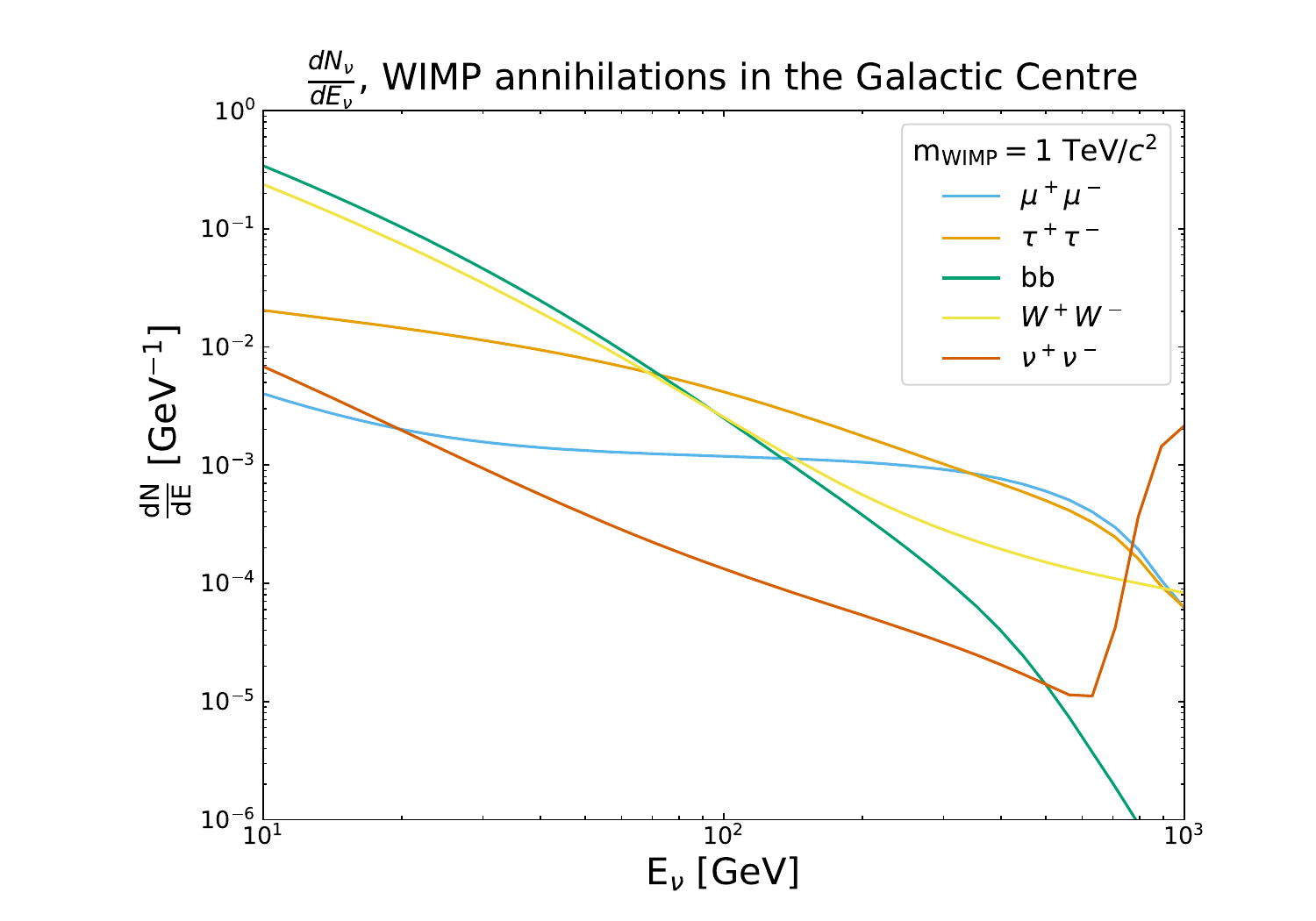}
    \caption{ Number of muon neutrinos produced in one WIMP-pair annihilation in the Galactic Centre as a function of the neutrino energy. The curves are shown for a WIMP mass of $m_{\mathrm{WIMP}} = 1 $ TeV$/c^2$ annihilating into muons, taus, b-quarks, W bosons and directly to neutrinos.}
    \label{fig:spectra}
\end{figure}

The expected neutrino flux at the Earth surface due to the annihilation of dark matter particles into secondary products can be expressed as

\begin{equation}
    \frac{d^2 \Phi}{d E d t} = \frac{\Gamma}{ 4 \pi D^2}  \frac{d N}{d E},
\end{equation}
\noindent where the parameter $\Gamma$ is the annihilation rate of WIMP particles, $D$ is the distance to the source and $\frac{dN}{dE}$ represents the number of neutrinos per unit energy emitted in one annihilation event. Neutrino spectra are computed for a set of WIMP masses from the GeV/$c^2$ to the 100 TeV/$c^2$ scale for five annihilation channels:

\begin{equation*}
    \mathrm{WIMP + WIMP} \rightarrow \mu^+ \mu^-, \tau^+ \tau^-, b \bar{b}, W^+W^-, \nu \bar{\nu}.
\end{equation*}
For each channel, a $100\%$ branching ratio is assumed, except for the neutrino channel, where a flavour-blind annihilation with a branching ratio of 1/3 to each neutrino flavour is adopted. These channels span the full range of realistic neutrino spectra, and the specific choice of hadronic/leptonic channels was made considering which annihilation modes yield the highest sensitivity in neutrino telescopes. For both the Galactic Centre and the Sun, the neutrino yields from the subsequent decays and emissions of the annihilation products are described using PYTHIA, a Monte Carlo event generator of high-energy physics collision events \cite{pythia}. The yields are implemented in the form of tables in the PPPC4DMID framework \cite{pppc} in the case of annihilations in the Galactic Centre. The neutrino yields from dark matter annihilations in the Sun are obtained with WimpSim \cite{wimpsim}, implemented within the DarkSUSY software package \cite{darksusy}. The determination of the neutrino spectra produced in both sources contains some common sources of uncertainty, such as the uncertainties
on showering and hadronisation in secondary SM processes and the treatment of electroweak corrections~\cite{charon}. Additional sources of uncertainty in the computation of neutrino spectra in the Sun are related to the transport of neutrinos through the solar medium: the treatment of the interactions and the uncertainties in the neutrino cross sections result in large differences in the predicted spectra in literature \cite{charon}. Neutrino oscillations are taken into account when propagating neutrinos from the source to the Earth \cite{nufit}: as the path they travel is sufficiently large, the energy and path length dependencies are taken as an average value for both sources. The number of muon neutrinos per unit energy emitted in the chain of processes resulting from one WIMP pair annihilation in the Galactic Centre is shown in Fig.~\ref{fig:spectra} for the five annihilation channels of interest, for a WIMP mass $m_{\mathrm{WIMP}} = 1  \: \mathrm{TeV} / c^2$. An identical number of antineutrinos is expected to be produced in the annihilation process.

\subsection{Neutrino flux from annihilations is the Galactic Centre}
 The WIMP annihilation rate, $\Gamma$, in the Galactic Centre depends on the spatial distribution of dark matter and the thermally-averaged cross section of the WIMP annihilation process, $\langle \sigma \upsilon \rangle$, which is the parameter being measured or constrained. The spatial distribution of WIMPs is expressed in terms of the Galactic density profile, which is estimated using the CLUMPY program \cite{clumpy}. In this analysis, the Navarro-Frenk-White (NFW) profile is used \cite{nfw}, with the local halo density set to $\rho_{\mathrm{local}} = 0.471 \: \mathrm{GeV}/c^2/\mathrm{cm}^3$ and the scale radius $r_s = 19.1 \: \mathrm{kpc}$, as used in the ANTARES publication \cite{antares_gc}. This density profile, expressed in terms of the angular distance from the centre of the galaxy, $\theta$, is integrated along the line of sight (l.o.s.) and through the solid angle that the object subtends in the sky, $\Delta \Omega$. The integrated quantity is often referred to as the J-factor. The neutrino flux at the surface of the Earth then reads as

\begin{equation}
    \frac{d^2 \Phi_{\nu}}{ dE_{\nu} dt} = \frac{1}{4\pi} \frac{1}{2} \frac{\langle \sigma \upsilon \rangle}{m_{
\mathrm{WIMP}}^{2}} \frac{dN_{\nu}}{dE_{\nu}} \int_{\Delta \Omega} {\int_{\mathrm{l.o.s.}}{\rho^{2}(\theta, l) dl \, d\Omega}}.
\label{eq:gc_flux}
\end{equation}
The factor $1/2$ appears as the WIMPs are assumed to be their own anti-particles and the squared WIMP mass, $m_{
\mathrm{WIMP}}^{2}$, as two WIMPs participate in the annihilation process. Flux expectations from 15 WIMP masses in the range from 500 GeV/$c^2$ to 100 TeV/$c^2$ are searched for in the Galactic Centre analysis. The density profile of dark matter in the Milky Way is largely unconstrained due to the difficulty of observing stellar objects close to the centre of the Galaxy. As a consequence, both cuspy and cored density profiles produce a similar level of agreement between theory and observation when fitted to the rotation curves of Milky Way objects \cite{mw_halo}. 
This factor is therefore the largest source of uncertainty in determining the neutrino flux. Cored density profiles, with constant densities within a few kpc of the Galactic Centre, produce a significantly more dispersed spatial signature, as the emission is less concentrated at the centre of the Galaxy. Consequently, neutrinos emitted from cuspy profiles will be more concentrated and easier to discriminate from the background.

\subsection{Neutrino flux from annihilations in the Sun}

The number of WIMPs captured within the Sun depends on the capture, annihilation and evaporation processes. Evaporation processes refer to ejections of WIMPs by hard scattering from nuclei, a process relevant for WIMP masses below a few GeV/$c^2$ \cite{darksusy}. This analysis probes masses above 10 GeV/$c^2$, for which evaporation is irrelevant and the equilibrium time for capture and annihilation processes is below the age of the Solar System \cite{darksusy}. Equilibrium between capture and annihilation processes implies $\Gamma = C_r/2 $, $C_r$ being the capture rate of dark matter in the Sun. The latter is related to the WIMP-nucleon scattering cross section, which can be spin-dependent or spin-independent. If the coupling between WIMPs and nucleons is a spin-dependent axial-vector coupling, WIMPs can only scatter off nuclei with net spin, predominantly hydrogen nuclei. If the coupling between WIMPs and nucleons is a spin-independent scalar coupling, the WIMPs can scatter off all the isotopes composing the Sun, therefore a model of the elemental solar abundances and their mass fractions must be adopted~\cite{darksusy}. Each target nucleus has a different mass, resulting in a different form factor suppression and a different kinematic suppression of the capture. All this can be accounted for with the following relation between the WIMP-nucleon cross section and the neutrino flux at the surface of the Earth:

\begin{equation}
    \sigma^{\mathrm{SD,SI}} = K^{\mathrm{SD,SI}} \Phi_{\nu+\bar{\nu}},
\label{eq:sun_limits_cs}
\end{equation}
where $\sigma^{\mathrm{SD,SI}}$ represents the spin-dependent/spin-independent WIMP-nucleon scattering cross section and $K$ is a conversion factor which contains information about the WIMP-WIMP annihilation channel, the mass of the WIMP, the different suppression factors and abundances of each element, the local density and velocity distributions of the WIMP halo surrounding the Sun, the Earth-Sun distance and the neutrino oscillation probabilities. The conversion factors are computed using the software package DarkSUSY. Three annihilation channels are considered when searching for dark matter from the Sun: $\tau^+ \tau^-$, $b \bar{b}$ and $W^+ W^-$. Annihilations into muons are excluded as muons lose too much energy in the core of the Sun before they produce a neutrino, whereas flux expectations for the neutrino channel were not available in DarkSUSY at the time of writing. For the Sun analysis, 36 masses in the range between 10 GeV/$c^2$ and 5 TeV/$c^2$ are considered.


\section{The KM3NeT detector} \label{sec:km3net}

\subsection{The KM3NeT detector and data sets}

The KM3NeT research infrastructure hosts two underwater Čerenkov detectors in the Mediterranean Sea, named ARCA and ORCA \cite{loi}. The operation principle relies on the detection of the Čerenkov light induced by ultra-relativistic charged particles propagating in the vicinity or inside the detector. The light is detected by three-dimensional arrays of digital optical modules (DOMs) \cite{dom}. Each DOM is composed of 31 photomultiplier tubes (PMTs) housed in a pressure-resistant glass sphere, along with the associated readout electronics and sensor devices for calibration and monitoring. The DOMs are grouped into vertical lines called detection units (DUs), each hosting 18 DOMs. 

PMT signal pulses exceeding  a tunable threshold voltage are digitised and their start time and pulse duration are recorded. Together with the PMT identification number these data form a {\it hit}. All digitised hits are transmitted to shore where they are filtered and processed by means of trigger algorithms which form events by selecting causally-connected hits.

The ARCA (Astroparticle Research with Cosmics in the Abyss) detector is located 100 km offshore the Sicilian coast near Portopalo di Capo Passero
(Italy) down to a depth of 3500~m. Its DOMs are vertically spaced by about 36 m on a DU and the DUs are horizontally spaced by about 90~m. The goal of ARCA is to detect astrophysical neutrinos at energies of hundreds of GeV up to PeV energies. The ORCA (Oscillation Research with Cosmics in the Abyss) detector is situated near the coast of Toulon, France, 40~km offshore and anchored to the seabed at 2500~m depth. The configuration of this detector is denser: DOMs are spaced vertically by about 9 m on a DU and the DUs are horizontally spaced by about 20~m. This is optimised for the detection of atmospheric neutrinos at GeV energies up to $\sim$100 GeV, with the main goal of measuring neutrino oscillation parameters and determining the neutrino mass ordering. The two different detector configurations allow for testing of a wide range of dark matter models, with WIMP masses ranging from a few GeV$/c^2$ up to the theoretical limit for thermally produced WIMPs at hundreds of TeV$/c^2$ \cite{unitarity}.

 The two detectors are currently under construction. The ARCA detector will comprise two blocks of 115 DUs, while ORCA will consist of one single block of 115 DUs. As of November 2024, the ARCA detector consists of 33 DUs and the number of DUs operating at ORCA is 24. The Galactic Centre analysis reported in this article was conducted on data collected by the ARCA detector with 8, 19 and 21 DUs, hereafter referred to as ARCA8, ARCA19 and ARCA21, respectively, between September 2021 and December 2022, for a total of 331 days. Regarding the Sun analysis, the data were taken with the ORCA configuration with 6 DUs, referred to as ORCA6, in operation between January 2020 and November 2021, for a total of 543 days. The period of operation and the effective livetime of each configuration are shown in Table \ref{livetime_table}. The effective livetime excludes periods with downtime due to technical issues or the installation of new DUs, calibration runs, periods with an extremely high bioluminescence or data taking runs that do not pass run quality criteria.

\begin{table}[]
\centering
\resizebox{\textwidth}{!}{
\begin{tabular}{|c|c|c|c|}
        \hline
        \textbf{Configuration} & \textbf{analysed period} & \textbf{effective livetime [days]} & \textbf{number of selected events} \\
        \hline
        ARCA8 & 26.09.2021 - 01.06.2022 & 210 & 647 \\
        \hline
        ARCA19 & 10.06.2022 - 12.09.2022 & 52 & 517 \\
        \hline
        ARCA21 & 22.09.2022 - 19.12.2022 & 69 & 1044 \\
        \hline
        ORCA6 & 26.01.2020 - 18.11.2021 & 543 & 2366 \\
        \hline
    \end{tabular}
}
\caption{The data taking period, the effective livetime and the number of selected events of each data set after the application of a quality run selection and an event selection as described in the text. The detector configuration is defined as the number of active detector units, the number ending the ARCA and ORCA acronyms.}
\label{livetime_table}
\end{table}

 \subsection{Event generation, propagation and reconstruction}

 The analysis methods are optimised on simulated events. A GENIE-based \cite{genie} code named gSeaGen \cite{gseagen} is used to generate neutrino interactions in water and the resulting flux of neutrinos at the detectors. The MUPAGE package \cite{mupage} is used to simulate the atmospheric muon flux at the detectors, produced in cosmic ray collisions in the atmosphere. The muon rate is calculated from parametric formulae according to the depth in sea \cite{parametric_formulae}. The propagation of the particles produced in neutrino interactions and the resulting Čerenkov light emission are handled by an internal software package.
Low energy events in ORCA are propagated with KM3Sim \cite{km3sim}, a package based on GEANT4 \cite{GEANT4} that traces the particle trajectories through the medium, generates Čerenkov photons and propagates the photons to the PMT surface. In the case of high energy events at ARCA, a GEANT4 simulation of the detector is computationally too demanding. Instead, the light reaching the PMT surfaces is computed using probability density functions (PDFs) of the arrival time of photons, which depend on the distance of the PMT to the particle trajectory, the particle energy and the incident angle of the light on the PMT surface. Absorption and scattering of light in water are taken into account in both detectors when photons are propagated. The simulation of the optical background due to the PMT dark current and the decays of $^{40}$K present in sea water, as well as the PMT response and readout, are handled by a dedicated KM3NeT software package. The background rates and the status and configuration of individual PMTs are inferred from the data in order to accurately simulate data taking conditions.

Neutrino events detected by KM3NeT consist of mainly deep inelastic scatterings of neutrinos on nucleons via the exchange of a $W^{\pm}$ boson, charged-current (CC) interactions, or via a $Z^0$ boson, neutral-current (NC) interactions. Muon neutrinos interacting via CC interactions produce muons that traverse the detector in a straight line whilst inducing the emission of a cone of Čerenkov light. The light is emitted at a well-defined, characteristic Čerenkov angle, which is $\sim 42^{\circ}$ for sea water at the relativistic energies observed by KM3NeT. These events are referred to as \textit{track-like}. Muons crossing the detector can pass through the entirety of the instrumented volume, leaving Čerenkov photon signals in DOMs around the muon trajectory. This allows for a precise measurement of the muon direction whose angular resolution improves with energy, reaching a sub-degree accuracy for energies higher than 1 TeV. Tau leptons produced in tau neutrino CC interactions give rise to track-like events when the tau lepton decays to a muon (branching ratio $\sim 17\%$) traversing the detector volume. Other types of neutrino interactions will produce electromagnetic or hadronic showers at the neutrino interaction vertex, resulting in a more isotropic photon distribution. This class of shower-like event topologies is not used in this work. 

 
 Reconstruction algorithms \cite{event_reconstruction} are used to obtain an estimate of the direction and energy of the events from the PMT hit patterns in the detector assuming a given event topology. For the track topology, the reconstruction begins with a linear prefit which provides a set of best-fit solutions for the particle direction given the measured hit times at the triggered PMTs. These solutions are then used as starting points for a likelihood maximisation, where the likelihood function is a product of the PDFs of the arrival time residuals at each PMT. The residuals are computed as the difference between the measured arrival time of the photons and the expected arrival time from the reconstructed track hypothesis. The PDFs account for Čerenkov emission and energy losses of the particle, scattering and absorption of the produced photons, the quantum efficiencies of the PMTs and the background emission from $^{40}$K decays. The starting time of the event, the interaction vertex and track direction are fitted by maximising the likelihood. The event energy is subsequently fitted after the most likely direction and interaction vertex of the event are found.
 
\subsection{Event selection}

The analyses reported in this work are conducted on track-like events, where the largest source of background are atmospheric muons. In order to reject them, only upgoing events that enter the atmosphere and traverse the Earth before arriving at the detector are considered. Additional selection cuts are applied in order to remove noise events, poorly reconstructed tracks and muons that could be mis-reconstructed as upgoing. Noise events are removed by requiring a minimum number of PMT hits in the event and a certain value for the maximised likelihood of the reconstructed track. 

\subsubsection{ORCA6 event selection}


Two more sets of cuts are applied for ORCA6 due to its smaller number of DUs (track reconstruction is more difficult) and its lower energy threshold (higher flux of atmospheric muons and neutrinos). The first set is based on the agreement between the recorded and the expected arrival time of the Čerenkov photons emitted along the muon track. A minimum number of hits compatible with the track hypothesis is required. The second set of cuts are containment cuts requiring the reconstructed vertex position to be inside the instrumented volume. The distance from the interaction vertex to the detector centre is required to be below 60 m. In addition, events with vertices in the upper 55 m of the detector are rejected. The veto on the upper layers is placed in order to further suppress the downgoing muon contamination. Further cuts on the  track likelihood and the estimated angular error in reconstruction are optimised in order to obtain the maximum sensitivity.

\subsubsection{ARCA8/19/21 event selection}

\begin{figure}[h!]
    \centering
    \begin{subfigure}{0.5\textwidth}
        \centering
        \includegraphics[width=\linewidth]{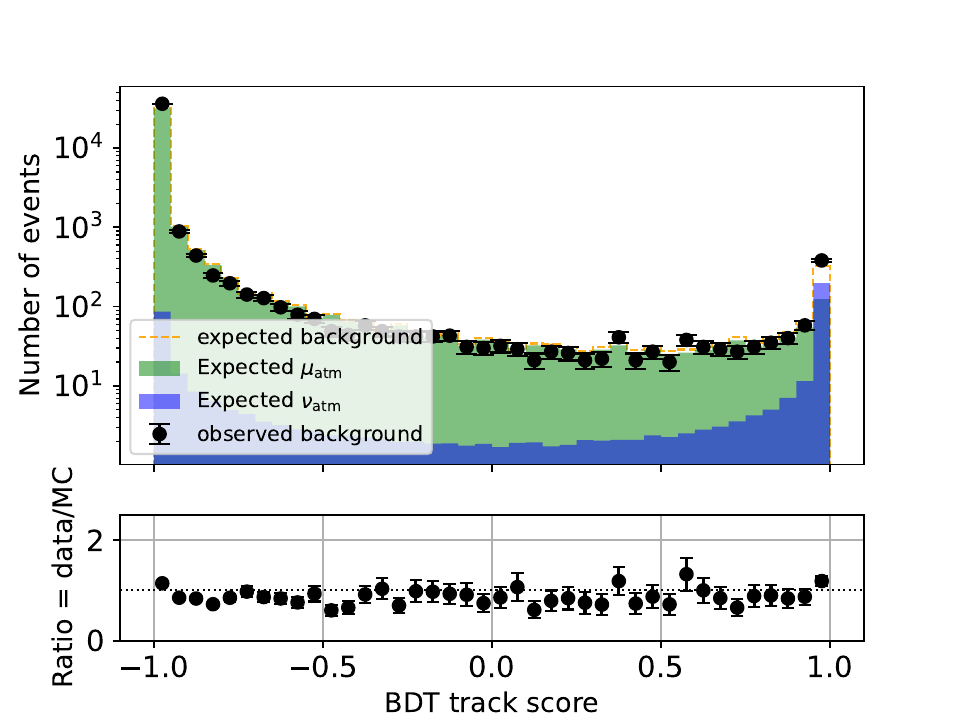}
        \label{subfig:figure_a19}
    \end{subfigure}%
    \hfill
    \begin{subfigure}{0.49\textwidth}
        \centering
        \includegraphics[width=\linewidth]{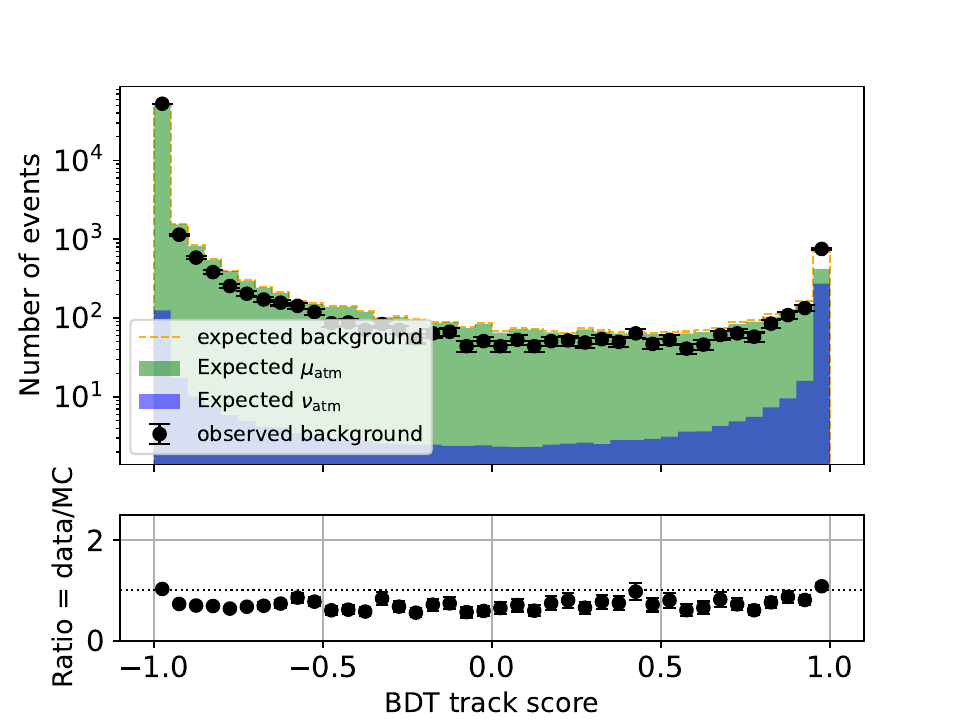}
        \label{subfig:figure_a21}
    \end{subfigure}
    
    \caption{BDT score distributions of data events and the MC background expectation prior to the application of the final event selection, shown for the ARCA19 detector (left) and ARCA21 (right). Black dots indicate data events, the atmospheric muon simulation is shown in green, the atmospheric neutrino simulation in blue and the dashed line indicates the total number of simulated background events. The ratio plot compares the number of data events in each bin to the total number of expected background events from simulations. The final selection places a cut on the BDT score at 0.95, for both the ARCA19 and ARCA21 detectors.}
    \label{fig:bdt_scores}

\end{figure}

Similarly to the ORCA6 event selection, the ARCA8 selection is based on cuts on reconstructed variables such as the track likelihood, the estimated track length and the angular error, which are optimised in order to obtain the best sensitivities.

Finally, for the ARCA19 and ARCA21 samples, a boosted decision tree (BDT) algorithm was trained in order to identify well-reconstructed tracks. The BDT features are 20 variables based on the track reconstruction algorithm. The variables with the highest importance include the reconstructed track length, the number of photoelectrons produced at the PMTs along the whole track and the reconstruction angular error estimate. Other variable inputs to the BDT include variables which describe how well the linear prefit stage of the reconstruction converges: the number of reconstructed linear prefit track solutions with upgoing/downgoing direction, the number of prefit solutions within one degree of the best fit direction, the maximum zenith difference between the prefit solutions. Variables which test the track/shower emission hypothesis more rigorously also help discriminate neutrinos from muons: the distance between the closest and furthest hit from the interaction vertex having a time residual below 15 nanoseconds, the number of such hits within 100 m of the vertex compared to the total number of such hits, the number of hits with a small time residual for both the track and shower hypothesis. The distribution of the BDT score of data events and the atmospheric background simulation is shown in Fig.~\ref{fig:bdt_scores}.

\begin{figure}[h!]
    \centering
\includegraphics[width=\linewidth]{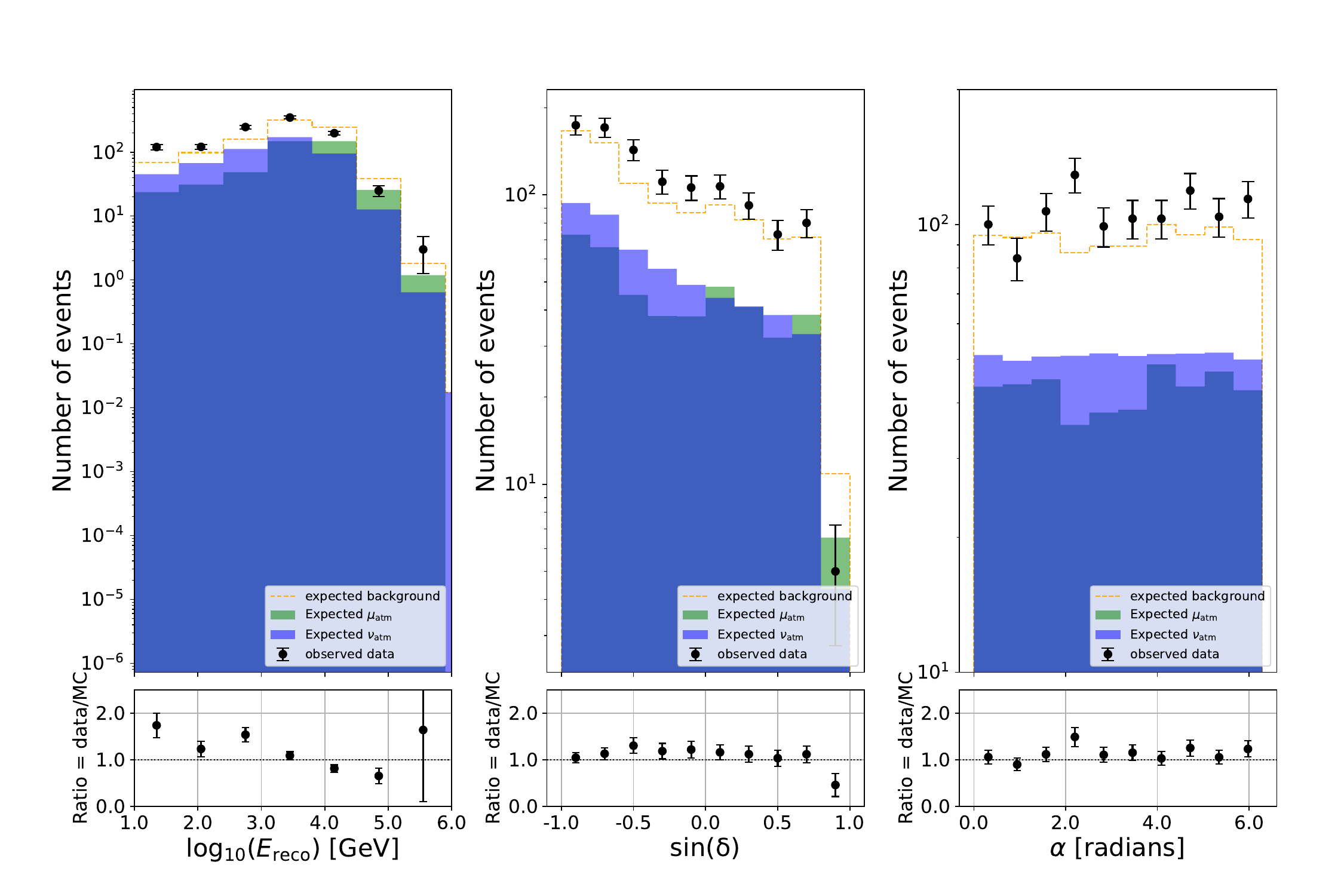}
    
    \caption{The distributions of the reconstructed energy (left), sine of the declination (middle) and the right ascension (right) in ARCA21 after the application of the event selection. The distributions of data events are shown as black dots, the expected distribution of atmospheric muons is shown in green and the simulation of atmospheric neutrinos is shown in blue, while the dashed line is the total MC expected background. The ratio plot shows the ratio between the number of data events and the total MC background in each bin.}
    \label{fig:data_mc_zenith_ereco}

\end{figure}

The BDT was trained with events satisfying certain conditions: the reconstructed track must contain a hit in at least two DOMs in the detector and the reconstructed angular error must be below one degree. The final cut on the BDT track score is optimised aiming at the increase in the signal efficiency and the reduction of the atmospheric muon contamination while ensuring a good agreement between data and simulations. The resulting muon contamination and the efficiencies of event selections applied to each ARCA configuration are shown in Table~\ref{efficiency_table}. The atmospheric background expectation obtained from MC simulations and the observed background distributions of the reconstructed energy and the equatorial coordinates can be seen in Fig.~\ref{fig:data_mc_zenith_ereco}. Distributions of the ARCA21 detector configuration are shown, as this is the configuration offering the most sensitivity to neutrinos from WIMP annihilations.

\begin{table}[]
\centering
\resizebox{\textwidth}{!}{
\begin{tabular}{|c|c|c|c|}
        \hline
        \textbf{Configuration} & \textbf{muon contamination} & \textbf{selection efficiency [1 TeV/$c^2$]} & \textbf{selection efficiency [100 TeV/$c^2$]} \\
        \hline
        ARCA8 & 60.2\% & 41.2\% & 49.2\% \\
        \hline
        ARCA19 & 27.2\% & 68.8\% & 77.2\%\\
        \hline
        ARCA21 & 43.3\% & 73.9\% & 80.7\% \\
        \hline
    \end{tabular}
}
\caption{The muon contamination and selection efficiencies for WIMP masses of 1 TeV/$c^2$ (column 3) and 100 TeV/$c^2$ (column 4) for the muon annihilation channel are shown for the three ARCA configurations analysed. The selection efficiency is defined as the ratio between the selected muon neutrino signal events after the application of the final selection and the number of muon neutrino signal events detected after the application of an upgoing, anti-noise selection only.}
\label{efficiency_table}
\end{table}

\section{Analysis Method} \label{sec:method}

This analysis aims to distinguish a cluster of dark matter signal events around a source centre (the signal hypothesis, H$_{1}$) from the null hypothesis (H$_{0}$), for which all the events originate from the atmospheric background, using an unbinned extended likelihood. The unbinned likelihood function is built from the PDFs of the signal and background hypotheses. The PDFs are two-dimensional distributions of the reconstructed neutrino directions and energies. Given the angular distance from the source centre and the event reconstructed energy, a probability to be signal or background is assigned to each event. For the search in the Galactic Centre, the halo density profile is used in combination with the detector angular response to model the spatial distribution of the dark matter-induced signal with respect to the source centre. As its size is smaller than the angular accuracy of the detector, the Sun is treated as a point source, so the spatial distribution of the signal is obtained from the angular response of the detector alone. The spectra of neutrinos coming from WIMP annihilations are used to reweigh simulated neutrino events and estimate the signal energy distribution at the detector. The background PDF is obtained from blinded data, assigning random right ascension coordinates to the events. The distribution of the declination of the events is used to characterise the background events in space for the  Galactic Centre analysis, whereas in the case of the Sun the source is moving, so the analysis is conducted in coordinates centred at the source location. The background is thus characterised by comparing the event locations to locations of the Sun at random times, differing from the event time. For both analyses, the reconstructed energy distribution of observed events is used to estimate the energy distribution of background events.

\subsection{Galactic Centre analysis method}

In the case of the Galactic Centre analysis, separate PDFs are produced for each ARCA configuration and the analysis is conducted on joint data sets including all events. The likelihood function used is given by
\begin{equation}
    \mathrm{log} ( \mathcal{L} ) = \sum_{\substack{d}} \sum_{i=0}^{N_{d}} \mathrm{log} [ f_{d} \cdot n_{\mathrm{sg}} \cdot \mathcal{S}^{d} (\alpha_i, E_i) + (N_{d} - f_{d} \cdot n_{\mathrm{sg}}) \cdot B^{d} (\alpha_i, E_i) ] - N_{t}.
    \label{likelihood_eq}
\end{equation}

\par \noindent The sum, using $d$ as index, is performed over the different ARCA configurations. $N_{d}$ is the total number of events present in configuration $d$; $\alpha_i$ and $E_i$ denote the angular distance from the source and the reconstructed energy of the event $i$; $\mathcal{S}^{d}$ and $B^{d}$ denote the signal and background probabilities obtained from PDFs for the event in the configuration $d$; $n_{\mathrm{sg}}$ and $N_{t}$ denote the number of signal events and the total number of events, summed for all configurations. The factor $ f_{d}$ quantifies the relative amount of signal detected in each detector configuration and it is given by the product of the signal acceptance and livetime of configuration $d$. The signal acceptances are computed as a convolution of the detector effective area, $A_{\mathrm{eff}_{d}} (\delta, E_{\nu})$, and the WIMP annihilation spectrum:
\begin{equation}
    \mathrm{Acc_d} = \frac{\int_{E_{\mathrm{th}}}^{m_{\mathrm{WIMP}}} A_{\mathrm{eff}_d} (E_{\nu}, \delta) \frac{d N }{d E} \space d E_{\nu}}{ \int_{E_{\mathrm{th}}}^{m_{\mathrm{WIMP}}} \frac{d N }{d E} \space d E_{\nu}} .
\end{equation}
The effective area is defined as the ratio between selected events in the simulations and the total simulated neutrino flux for a given data taking period. It is integrated from the minimum threshold energy that can be reconstructed in the detector, $E_{\mathrm{th}}$, up to the WIMP mass. This threshold energy is equal to 100 GeV for the ARCA detector, and 1 GeV for the ORCA detector. The effective area is also dependent on the event declination, $\delta$: for the Galactic Centre, the effective area computed at the declination bin corresponding to the source location is used in the acceptance calculation.

\subsection{Sun analysis method}
 
\begin{figure}
    \centering
    \includegraphics[width=0.8\textwidth]{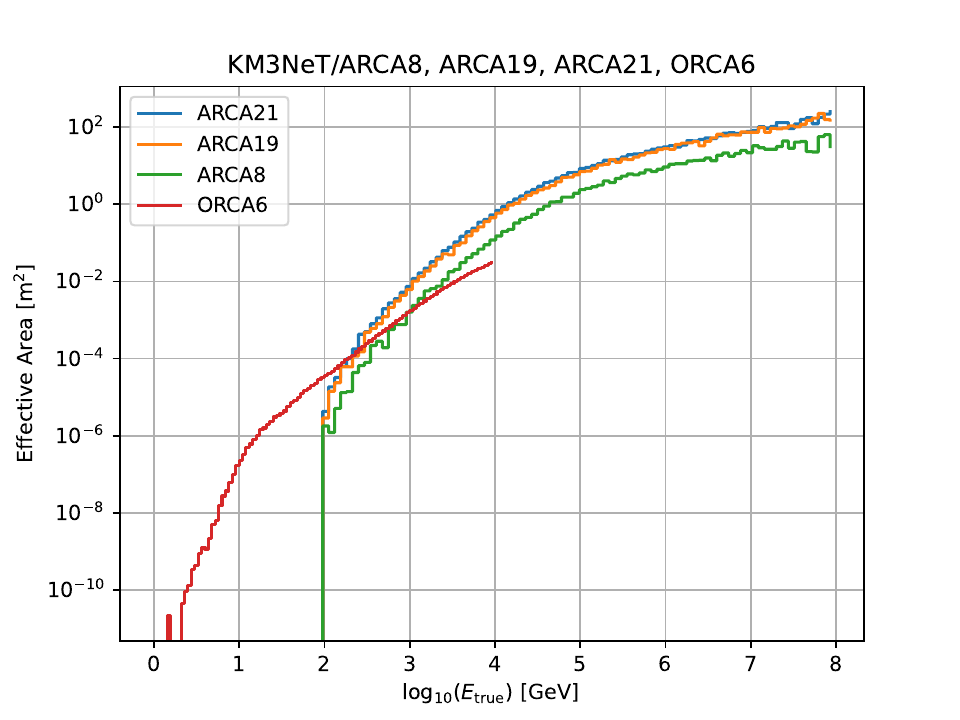}
    \caption{The effective area as a function of the true energy for muon neutrinos of the ARCA8, ARCA19 and ARCA21 detectors at the Galactic Centre declination, and the ORCA6 detector weighting declination bins according to the fractional time the source spends in each bin. The effective areas are computed after the application of the event selection.}
    \label{fig:aeff}
\end{figure}

The method for the Sun analysis is similar: the likelihood function for the ORCA data set, for which only one configuration is present, is obtained by setting $f_d=1$ in Eq. \ref{likelihood_eq} and removing the summation over configurations: 
\begin{equation}
    \mathrm{log}(\mathcal{L}) =  \sum_{i=0}^{N_{t}} \mathrm{log} [ n_{\mathrm{sg}} \cdot \mathcal{S} (\alpha_i, E_i) + (N_{t} - n_{\mathrm{sg}}) \cdot \mathcal{B} (\alpha_i, E_i) ] - N_{t}.
    \label{likelihood_orca}
\end{equation}
The effective area is computed by weigthing effective areas in different declination bins according to the fractional amount of time the Sun spends in that bin. A weighted average of the effective areas in each of the bins is then used to compute the signal acceptance. The resulting effective areas of the ARCA8, ARCA19 and ARCA21 at the source declination is shown in Fig.~\ref{fig:aeff}, along with the effective area for ORCA6, computed with the above described method of weighting declination bins.

\subsection{Limit on the number of signal events}

For both the Galactic Centre and the Sun analyses, the significance of a signal cluster is evaluated with the test statistic
\begin{equation}
    \mathrm{TS} = \frac{\mathcal{L}(n_{\mathrm{sg}} ^ {\mathrm{max} } ) }{\mathcal{L}(n_{\mathrm{sg}}=0)},
\end{equation}
\noindent 
where $n_{\mathrm{sg}} ^ {\mathrm{max} }$ is the number of signal events that maximises the likelihood defined in Eq. \ref{likelihood_eq} and  Eq. \ref{likelihood_orca} and $\mathcal{L}(n_{\mathrm{sg}}=0)$ denotes the likelihood for a data set consisting of solely background events. The TS distributions for the signal and null hypotheses are built generating pseudo-experiments using the corresponding PDFs. ``Mock'' sky maps with a number of injected signal events varying between zero and 50 are generated. A total of $10^5$ mock sky maps containing only background events were generated, whereas $10^4$ sky maps were produced for each given number of signal events. A different TS distribution is obtained for each number of injected signal events.
\noindent
A convolution of the TS distribution with a Poisson function, $\mathcal{P}$, with mean $\mu$, is applied to account for Poisson fluctuations of the injected signal in each pseudo-experiment, as follows:
\begin{equation}
    P(\mathrm{TS}(\mu)) = \Sigma_{n_{\mathrm{sg}}} P(\mathrm{TS}(n_{\mathrm{sg}} )) \times \mathcal{P}(n_{\mathrm{sg}}, \mu).
\end{equation} 

The number of signal events detected in each pseudo-experiment is subject to uncertainties due to the limited accuracy in measuring the optical water properties in the detector, such as the light absorption length, and due to uncertainties in the PMT efficiencies. The uncertainties are accounted for by applying a $30\%$ and $15\%$ Gaussian smearing to the TS distributions, for ARCA and ORCA respectively \cite{fraf_icrc}. This value was obtained by simulating events with a modified light absorption length in water, as the uncertainty in this parameter exerts the largest influence on the number of detected events \cite{fraf_icrc}. These systematic uncertainties are applied independently of the event energy. The uncertainty is larger in ARCA because the distance between the detector components is significantly larger. Uncertainties on the light absorption length will therefore produce a larger variation in the event rates in ARCA where the light has to travel a larger distance.

The Neyman approach \cite{neyman} is followed to obtain sensitivities and upper limits on the number of signal events. The sensitivity on the number of events, $\bar{\mu}_{90}$, is defined as the averaged 90$\%$ confidence level (CL) upper limit for a measurement that coincides with the median of the background TS distribution. The 90\% CL upper limits  on the detected signal events, $\mu_{90}$, are computed from the TS of the unblinded data. If this TS is below the median background, the limit is set to the sensitivity.

In the absence of a signal, the limits in the number of signal events, $\mu_{90}$, are converted into  limits on the integrated flux with the following equation:

\begin{equation}
    \Phi_{\nu + \bar{\nu}}^{90} = \frac{\mu_{90}}{ \sum_{d} T_d \mathrm{Acc}_{d}  } ,
\end{equation}

\noindent
where $T_{d}$ and $\mathrm{Acc}_d$ are the livetime and the acceptance to signal events of data set $d$. Equations \ref{eq:gc_flux} and \ref{eq:sun_limits_cs} are then used to convert the flux upper limits into cross section upper limits in the case of the Galactic Centre and the Sun. Following a blind approach, the event selection criteria to be applied to the data are first optimised for each annihilation channel/WIMP mass combination to attain the best flux sensitivities before looking at the data.

\section{Results}

\subsection{Searches in the Galactic Centre} \label{sec:results_gc}

\begin{figure}[h!]
    \centering
    \includegraphics[width=0.8\textwidth]{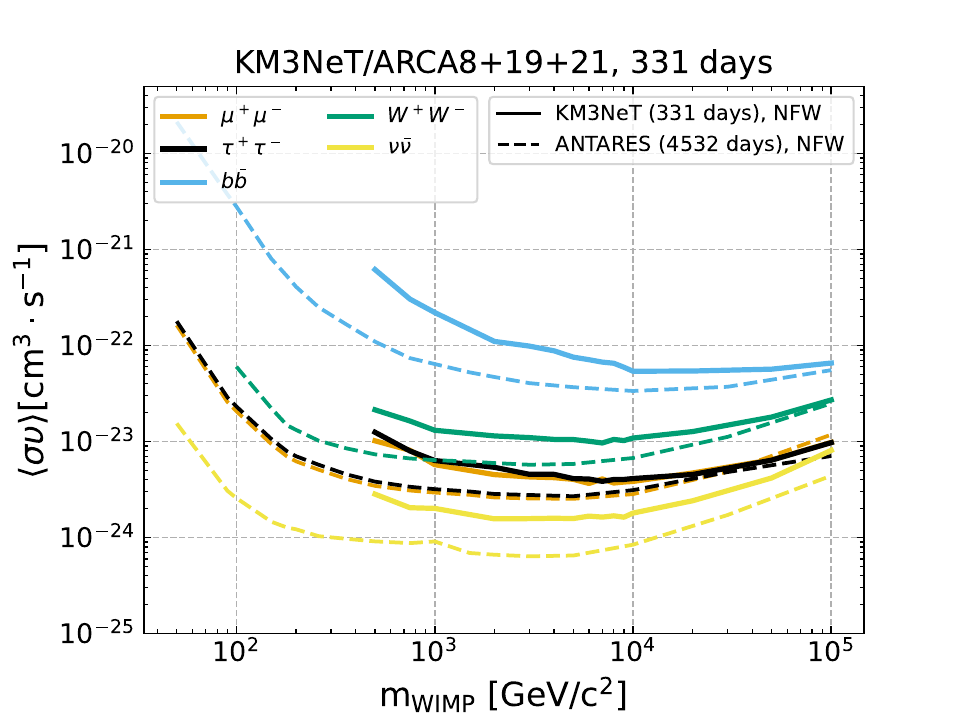}
    \caption{$90\%$ CL upper limits on the thermally-averaged WIMP annihilation cross section, $\langle\sigma \upsilon\rangle$, as a function of the WIMP mass for each of the five annihilation channels, obtained with the ARCA8-21 data set (full lines) and with the ANTARES 4532 day data set (dashed lines)~\cite{antares_gc}.}

    \label{fig:antares_results}
\end{figure}

\begin{figure}[h!]
    \centering
    \includegraphics[width=0.8\textwidth]{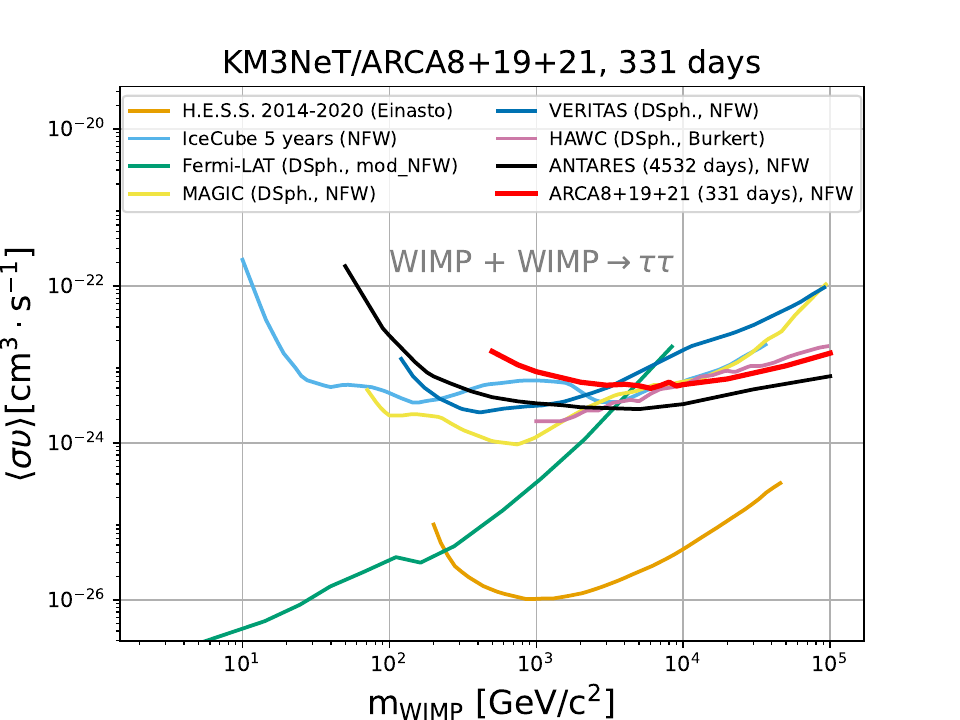}   
    \caption{$90\%$ CL upper limits on the thermally-averaged WIMP annihilation cross section, $\langle \sigma \upsilon \rangle$, as a function of the WIMP mass for the $\tau^+ \tau^-$ annihilation channel obtained with the ARCA8-21 data set along with results obtained by other experiments \cite{antares_gc, hess, IC, fermiNew, magic, veritas, hawc}.}

    \label{fig:gc_other_experiments}
\end{figure}

The data sets of the ARCA8-21 detector configurations were analysed in search of a WIMP annihilation signal for WIMP masses in the range 500 GeV$/c^{2} - 100$ TeV/$c^{2}$.
The TS of the data is found to be compatible with the background hypothesis for all combinations of WIMP masses and annihilation channels.
Bounds on the thermally-averaged WIMP annihilation cross section, $\langle\sigma \upsilon \rangle$, are placed using Eq. \ref{eq:gc_flux}. As previously mentioned, we use the NFW dark matter halo profile~\cite{nfw}. The 90\% CL limits on $\langle\sigma \upsilon \rangle$  as a function of the WIMP mass obtained using ARCA8-21 data are shown for the five channels under investigation in Fig.~\ref{fig:antares_results}. For comparison, the limits on the same quantity obtained using the full ANTARES data set are also shown \cite{antares_gc}.
The results for the $\tau^+ \tau^- $ channel are compared to other indirect searches in Fig.~\ref{fig:gc_other_experiments}: the H.E.S.S. inner galaxy survey \cite{hess}, the IceCube Galactic Centre and halo search \cite{IC}, the Fermi-LAT, MAGIC, VERITAS and HAWC dark matter searches in dwarf spheroidal galaxies \cite{fermiNew, magic,veritas, hawc}.  All constraints on the annihilation cross section derived from the Galactic Centre strongly depend on the chosen parametrisation of the halo density profile. Other halo density profiles used by the experiments are those of Einasto~\cite{einasto} and Burkert~\cite{burkert}. The H.E.S.S. survey of the Galactic Centre region \cite{hess} provides the most stringent limit on the WIMP pair-annihilation cross section for WIMP masses above 200 GeV$/c^2$, due to the length of observation, source visibility and the choice of a cuspy Einasto density profile. Previous ANTARES publications have shown that the limits could change up to an order of magnitude, depending on the choice of density profile of the Milky Way \cite{ANTARES:oldlimits}. The IceCube result \cite{IC} is obtained with a data set recorded with the DeepCore strings, using the neighbouring strings of IceCube as a veto for atmospheric muons, as the source is located above the horizon, where the atmospheric muon contamination is large. This limits the sensitivity to WIMP annihilations at higher WIMP masses, beyond the TeV scale. The targets of $\gamma$-ray observatories in the Northern Hemisphere, dwarf spheroidal galaxies, are smaller dark matter reservoirs but the $\gamma$-ray background emission from these objects is expected to be negligible, resulting in competitive limits. Results obtained from these targets also depend on the chosen density profile, although the effect is smaller. Due to their smaller dimensions with respect to the Galactic Center region they can be considered as point-like, so the emission profile is independent of the chosen density distribution. Only the J-factor, the total integral of the squared mass density shown in Eq.~\ref{eq:gc_flux}, is affected. Despite using limited data sets in partial configurations, the present ARCA results are already highly competitive, thanks to the location of KM3NeT in the Northern Hemisphere, ideally positioned to observe neutrinos coming from the Galactic Centre.

\subsection{Searches in the Sun} \label{sec:results_Sun}

\begin{figure}[h!]
    \centering
    \begin{subfigure}{0.49\textwidth}
        \centering
        \includegraphics[width=\linewidth]{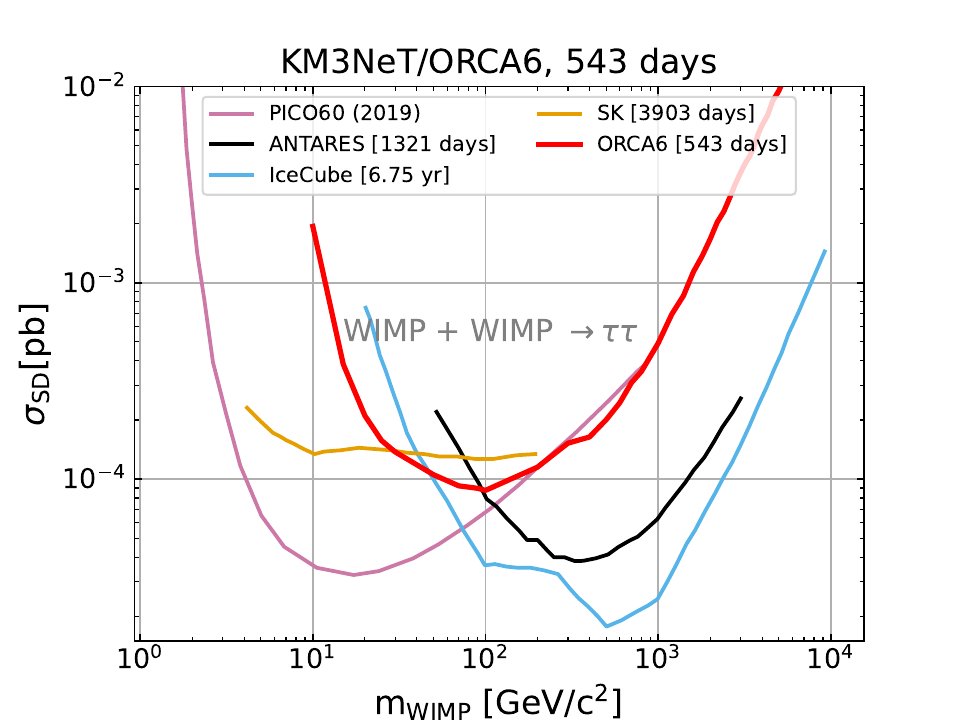}
        \label{subfig:figure2}
    \end{subfigure}%
    \hfill
    \begin{subfigure}{0.49\textwidth}
        \centering
        \includegraphics[width=\linewidth]{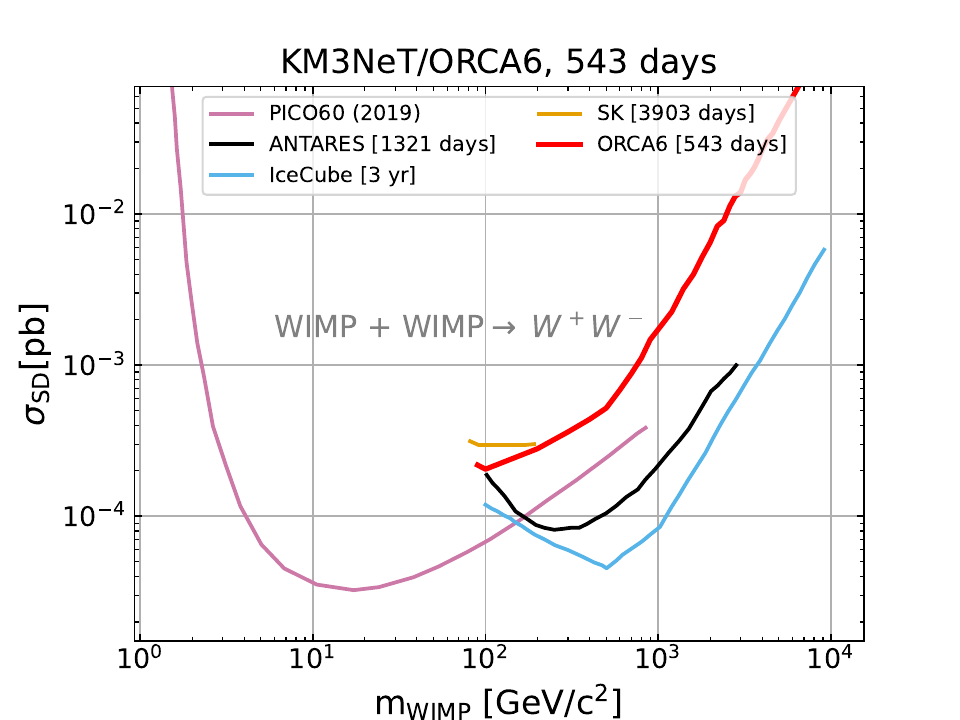}
        \label{subfig:figure3}
    \end{subfigure}
    
    \caption{$90\%$ CL upper limits on the spin-dependent WIMP-nucleon cross section as a function of the WIMP mass for the $\tau^+ \tau^-$ (left) and $W^+ W^-$ (right) annihilation channels. The red lines show the results obtained in this analysis, whereas the other lines show the upper limits obtained by IceCube \cite{IceCube_SUN, IceCube_sun_2}, ANTARES \cite{ANTARES_SUN} and Super-Kamiokande \cite{Super-Kamiokande_SUN}. The PICO-60 limit \cite{PICO_SUN} is obtained from the direct search of WIMP-nucleon scatterings and as such is independent of the annihilation channel considered.}
    \label{fig:Sun_cross_section_SD}
\end{figure}

\begin{figure}[h!]
      \centering
    \begin{subfigure}{0.49\textwidth}
        \centering
        \includegraphics[width=\linewidth]{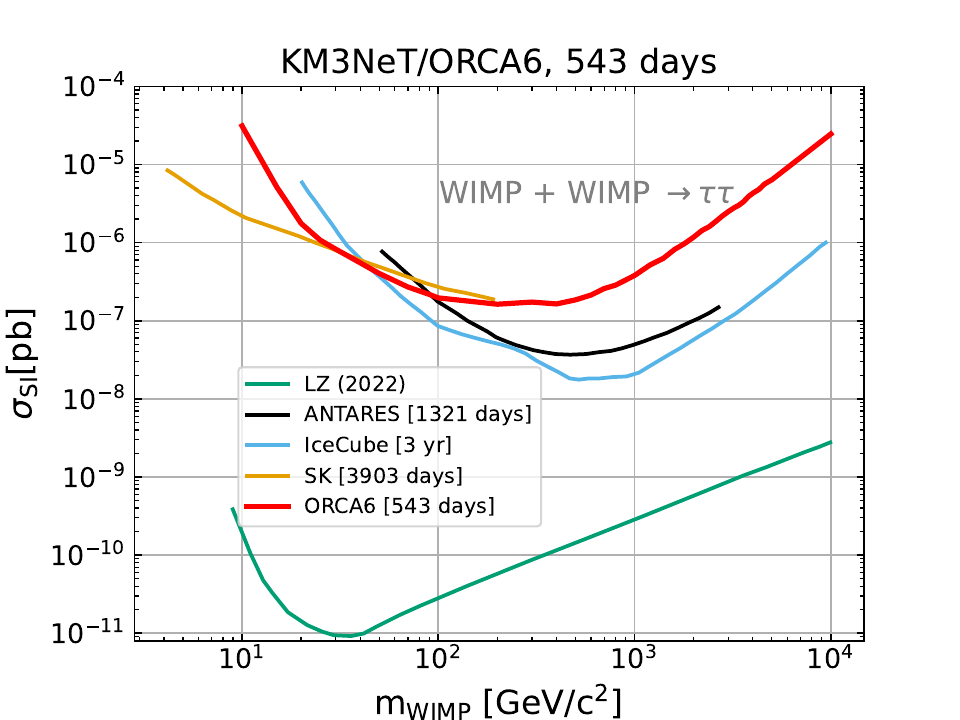}
        \label{subfig:figure2}
    \end{subfigure}%
    \hfill
    \begin{subfigure}{0.49\textwidth}
        \centering
        \includegraphics[width=\linewidth]{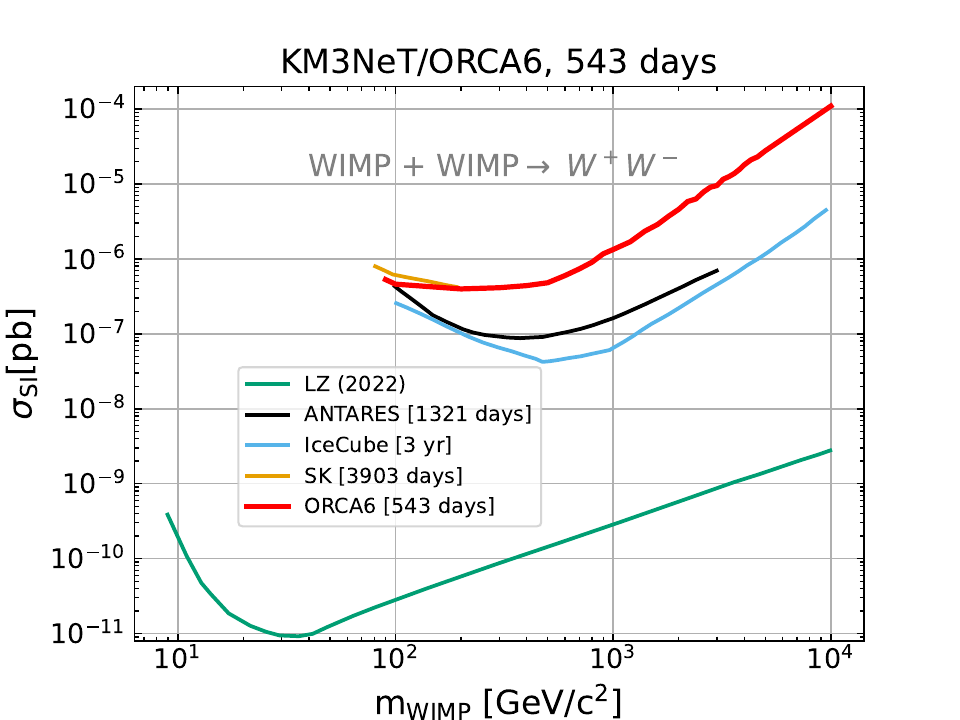}
        \label{subfig:figure3}
    \end{subfigure}
    \caption{$90\%$ CL upper limits on the spin-independent WIMP-nucleon cross section as a function of the WIMP mass for the  $\tau^+ \tau^-$ (left) and $W^+ W^-$ (right) annihilation channels. The red lines show the results obtained in this analysis, whereas the other lines show the upper limits obtained by IceCube \cite{IceCube_SUN, IceCube_sun_2}, ANTARES \cite{ANTARES_SUN} and Super-Kamiokande \cite{Super-Kamiokande_SUN}. The LZ \cite{LZ:2022lsv} limit is obtained from a direct search and as such is independent of the annihilation channel.}
    \label{fig:Sun_cross_section_SI}
\end{figure}

The ORCA6 data have been analysed in the search for dark matter in the Sun, considering WIMP masses in the range 10 GeV$/c^{2} - 10 $ TeV$/c^{2}$. As in the ARCA8-21 sample, the TS obtained for this data set is compatible with the background hypothesis for all combinations of WIMP masses and annihilation channels. In particular, the TS of the data is found to be below the median of the background-only TS distribution for every test case. Consequently, the limit on the neutrino flux is set to be equal to the sensitivity. Limits to the spin-dependent and the spin-independent cross sections are obtained through Equation \ref{eq:sun_limits_cs} and are shown in Figs. \ref{fig:Sun_cross_section_SD} and \ref{fig:Sun_cross_section_SI} respectively, where they are compared to limits from other indirect and direct searches. The ORCA6 results are already competitive when compared to those from other indirect detection experiments, which all have a significantly larger livetime. The upper limits obtained in direct detection experiments are independent of the WIMP annihilation channel, as the experiments attempt to detect the WIMP-nucleon scattering process directly. Searches for dark matter in the Sun in indirect detection experiments aim to observe the WIMP annihilation products and rely on the assumption of equilibrium between capture and annihilation in the Sun to place limits on the WIMP-nucleon scattering cross section. The results obtained in direct detection experiments are more stringent in the case of spin-independent WIMP-nucleon scattering. Indeed, the cross section for the spin-independent interaction grows with the nucleon mass, unlike the spin-dependent case, where the cross section scales with the nuclear spin. Direct detection experiments typically use a medium with heavy nuclear targets, such as liquid xenon, and can therefore set very stringent spin-independent cross section limits. On the other hand, in the case of spin-dependent WIMP-nucleon scattering, indirect detection experiments, including ORCA, have the opportunity to surpass current limits and improve the constraints on the WIMP-nucleon coupling.


\section{Summary} \label{sec:summary}

The first results on indirect searches for dark matter annihilation signatures with the KM3NeT neutrino telescopes have been presented. Due to their different configurations, the ARCA and ORCA detectors can cover a wide range of WIMP masses. The two telescopes, still in their construction phase, are already setting competitive limits on the dark matter coupling to the Standard Model. WIMP-nucleon cross section limits obtained with ORCA6 are surpassing their predecessor, ANTARES, at low WIMP masses, owing to the higher density of detector components and lower energy threshold. The annihilation cross section limits obtained with one year of livetime with ARCA8-21 are comparable, although less stringent, to those obtained with the full ANTARES data set, due to improved light detection technology and event reconstruction and selection methods. Follow-up searches with currently deployed and future larger detector configurations will push the boundary of dark matter searches with neutrino telescopes.

\section{Acknowledgements} The authors acknowledge the financial support of:
KM3NeT-INFRADEV2 project, funded by the European Union Horizon Europe Research and Innovation Programme under grant agreement No 101079679;
Funds for Scientific Research (FRS-FNRS), Francqui foundation, BAEF foundation;
Czech Science Foundation (GAČR 24-12702S);
Agence Nationale de la Recherche (contract ANR-15-CE31-0020), Centre National de la Recherche Scientifique (CNRS), Commission Europ\'eenne (FEDER fund and Marie Curie Program), LabEx UnivEarthS (ANR-10-LABX-0023 and ANR-18-IDEX-0001), Paris \^Ile-de-France Region, Normandy Region (Alpha, Blue-waves and Neptune), France,
For the CPER The Provence-Alpes-Côte d'Azur Delegation for Research and Innovation (DRARI), the Provence-Alpes-Côte d'Azur region, the Bouches-du-Rhône Departmental Council, the Metropolis of Aix-Marseille Provence and the City of Marseille through the CPER 2021-2027 NEUMED project,
The CNRS Institut National de Physique Nucléaire et de Physique des Particules (IN2P3); 
Shota Rustaveli National Science Foundation of Georgia (SRNSFG, FR-22-13708), Georgia;
This work is part of the MuSES project which has received funding from the European Research Council (ERC) under the European Union’s Horizon 2020 Research and Innovation Programme (grant agreement No 101142396).
The General Secretariat of Research and Innovation (GSRI), Greece;
Istituto Nazionale di Fisica Nucleare (INFN) and Ministero dell’Universit{\`a} e della Ricerca (MUR), through PRIN 2022 program (Grant PANTHEON 2022E2J4RK, Next Generation EU) and PON R\&I program (Avviso n. 424 del 28 febbraio 2018, Progetto PACK-PIR01 00021), Italy; IDMAR project Po-Fesr Sicilian Region az. 1.5.1; A. De Benedittis, W. Idrissi Ibnsalih, M. Bendahman, A. Nayerhoda, G. Papalashvili, I. C. Rea, A. Simonelli have been supported by the Italian Ministero dell'Universit{\`a} e della Ricerca (MUR), Progetto CIR01 00021 (Avviso n. 2595 del 24 dicembre 2019); KM3NeT4RR MUR Project National Recovery and Resilience Plan (NRRP), Mission 4 Component 2 Investment 3.1, Funded by the European Union – NextGenerationEU,CUP I57G21000040001, Concession Decree MUR No. n. Prot. 123 del 21/06/2022;
Ministry of Higher Education, Scientific Research and Innovation, Morocco, and the Arab Fund for Economic and Social Development, Kuwait;
Nederlandse organisatie voor Wetenschappelijk Onderzoek (NWO), the Netherlands;
Ministry of Research, Innovation and Digitalisation, Romania;
Slovak Research and Development Agency under Contract No. APVV-22-0413; Ministry of Education, Research, Development and Youth of the Slovak Republic;
MCIN for PID2021-124591NB-C41, -C42, -C43 and PDC2023-145913-I00 funded by MCIN/AEI/10.13039/501100011033 and by “ERDF A way of making Europe”, for ASFAE/2022/014 and ASFAE/2022 /023 with funding from the EU NextGenerationEU (PRTR-C17.I01) and Generalitat Valenciana, for Grant AST22\_6.2 with funding from Consejer\'{\i}a de Universidad, Investigaci\'on e Innovaci\'on and Gobierno de Espa\~na and European Union - NextGenerationEU, for CSIC-INFRA23013 and for CNS2023-144099, Generalitat Valenciana for CIDEGENT/2018/034, /2019/043, /2020/049, /2021/23, for CIDEIG/2023/20, for CIPROM/2023/51 and for GRISOLIAP/2021/192 and EU for MSC/101025085, Spain;
Khalifa University internal grants (ESIG-2023-008 and RIG-2023-070), United Arab Emirates;
The European Union's Horizon 2020 Research and Innovation Programme (ChETEC-INFRA - Project no. 101008324)

\clearpage

\bibliographystyle{JHEP} 
\bibliography{main} 

\end{document}